\documentstyle[aps,prl,multicol]{revtex}

\begin{document}
\input psfig.sty

\title
{\bf Parity Violation in Elastic Electron-Proton Scattering \\
and  the Proton's Strange Magnetic Form Factor}

\author{ 
D. T. Spayde$^{3}$, T. Averett\footnote{Present
address:  Department of Physics, College of William and Mary,
Williamsburg, VA 23187, USA}$^{1}$
 D. Barkhuff$^{4}$, D. H. Beck$^{2}$, E. J. Beise$^{3}$,  C. Benson$^{2}$,
 H. Breuer$^{3}$,
 R. Carr$^{1}$,  S. Covrig$^{1}$, J. DelCorso$^{5}$,
 G. Dodson$^{4}$, K. Dow$^{4}$, C. Eppstein$^{1}$,
M. Farkhondeh$^{4}$, 
B. W. Filippone$^{1}$,  P.Frazier$^{1}$, R. Hasty$^{2}$,
T. M. Ito$^{1}$, C. E. Jones$^{1}$,
W. Korsch$^{6}$, 
S. Kowalski$^{4}$, P. Lee$^{1}$, E. Maneva$^{1}$, K. McCarty$^{1}$,
R. D. McKeown$^{1}$, J. Mikell$^{2}$, B. Mueller$^{7}$, P. Naik$^{2}$,
 M. Pitt$^{5}$, J. Ritter$^{2}$,
V. Savu$^{1}$, M. Sullivan$^{1}$, R. Tieulent$^{3}$, E. Tsentalovich$^{4}$,
S. P. Wells$^{8}$, B. Yang$^{4}$, and T. Zwart$^{4}$
}

\address{
$^{1}$ Kellogg Radiation Laboratory, California Institute of Technology
Pasadena, CA 91125, USA \\
$^{2}$ Department of Physics, University of Illinois at Urbana-Champaign, Urbana, Illinois
61801 \\
$^{3}$ Department of Physics, University of Maryland, College Park, Maryland 20742 \\
$^{4}$ Bates Linear Accelerator Center, Laboratory for Nuclear Science
and Department of Physics,\\
Massachusetts Institute of Technology, Cambridge, Massachusetts 02139 \\
$^{5}$ Department of 
Physics, Virginia Polytechnic Institute and State University, Blacksburg, 
VA 24061-0435  \\
$^{6}$ Department of 
Physics and
Astronomy, University of Kentucky, Lexington, KY  40506  \\
$^{7}$ Physics Division, Argonne National Laboratory, Argonne, IL 60439, USA\\
$^{8}$  Department of Physics, Louisiana Tech University, Ruston, LA 71272, 
USA  \\
}

\maketitle
\vskip 10pt
\centerline{(SAMPLE Collaboration)}

\begin{abstract}
We report a new measurement of the  
parity-violating asymmetry in elastic
electron scattering from the proton at backward scattering
angles. This asymmetry is sensitive to the strange magnetic form factor
of the proton as well as electroweak axial radiative corrections.
The new measurement of $A=-4.92 \pm 0.61 \pm 0.73$ ppm provides a 
significant constraint on these quantities. The implications for 
the strange magnetic form factor
are discussed in the context of theoretical estimates 
for the axial corrections.
\end{abstract}

\begin{multicols}{2}[]
\narrowtext
\vfill
\eject

The anomalous magnetic moments of the neutron and proton are important
clues to their internal quark structure. Since the first measurement
of the proton's magnetic moment in 1933, our empirical knowledge of the 
electromagnetic structure of the nucleon has been greatly improved through
detailed measurements of the electric and magnetic form factors and
their momentum transfer ($Q^2$) dependence. 
Nevertheless, we still lack a quantitative theoretical
understanding of these properties (including the magnetic moments) and
additional experimental information is crucial in our effort to
understand the internal structure of the nucleons.

Whereas the normal magnetic moment corresponds to the magnetic coupling
to the photon, the weak magnetic moment represents the analogous coupling
to the $Z$ boson, equally fundamental and 
just as important as the electromagnetic moment.
The weak magnetic form factor provides unambiguous new information about the
quark flavor structure of the nucleon and 
enables a complete
decomposition of the proton's magnetic structure into the 
contributions from different quark flavors (up, down,
and strange)\cite{Kap88}. To lowest order,
the neutral weak magnetic form factor of the proton, $G_M^Z$, is related to 
nucleon 
electromagnetic form factors and a contribution from strange 
quarks:\footnote{This definition of $G_M^Z$ differs by a factor of 
4 from that used in
ref.~\cite{Muel97} in order to conform with more standard 
notation in the literature. The
definition of $G_M^s$ is the same as in ref.~\cite{Muel97}.}
\begin{eqnarray}  
G_M^Z =  (G_M^p-G_M^n) - 4 \sin^2\theta_W \> G_M^p - G_M^s
\end{eqnarray}  
where $G_M^p$ and $G_M^n$ are the (electromagnetic) nucleon magnetic form
factors, and $\theta_W$ is the weak mixing angle. (Electroweak 
radiative corrections to this expression have been computed in 
ref.~\cite{Mus90}.)
Thus the measurement of $G_M^Z$ provides unique access to the strange
quark-antiquark ``sea'' and its role in the basic electromagnetic structure
of the nucleons at low energies. $G_M^Z$ can be determined via 
parity-violating effects in elastic 
electron-proton  scattering\cite{BMcK89}. 

In this Letter, we report a new
measurement of the parity-violating asymmetry with sufficient precision
to provide the first meaningful information on the strange magnetic 
form factor, $G_M^s$. In comparison to our previous results~\cite{Muel97},
this measurement has involved both improved monitoring and control 
of systematic errors as well as improved statistical precision.

As previously discussed\cite{BMcK89}, 
the parity-violating asymmetry for elastic scattering of right-
vs.~left- handed electrons from nucleons at backward scattering angles
is quite sensitive to $G_M^Z$. The SAMPLE experiment measured 
the parity-violating
asymmetry in the elastic scattering of 200 MeV polarized electrons at 
backward angles with an average $Q^2 \simeq
0.1$(GeV/c)${}^2$. For $G_M^s=0$, the expected asymmetry in 
the SAMPLE experiment is about $-7\times 10^{-6}$ or 
-7 ppm, and the asymmetry depends linearly on $G_M^s$.
The neutral weak axial form factor $G_A^Z$ contributes about 20\% 
to the asymmetry in our experiment. 
In parity-violating electron scattering $G_A^Z$  
is modified by a substantial electroweak radiative correction. 
The corrections were estimated in~\cite{Mus90}, 
but there is considerable uncertainty in the calculations. 
The uncertainty in these
radiative corrections substantially limits our ability to determine
$G_M^s$, as will be discussed below.

The SAMPLE experiment was performed at the MIT/Bates Linear 
Accelerator Center using a 200 MeV polarized electron beam 
incident on a liquid hydrogen target. The scattered electrons
were detected in a large solid angle ($\sim 1.5$ sr) 
air {\v C}erenkov detector.
The detector consists of 10 large mirrors, each with ellipsoidal curvature
to focus the {\v C}erenkov light onto 
one of ten shielded photomultiplier tubes. 
A remotely controlled light
shutter can cover each photomultiplier tube for background
measurements. Typically one fourth
of the data was taken
with shutters closed to monitor this background.
As described in ref.~\cite{Muel97}, 
the {\v C}erenkov detector signals were studied at low beam currents to
determine the composition of the signal and the fraction of light 
due to elastic scattering (factors that scaled the individual mirror
asymmetries by typically 1.8, depending upon the
mirror, and that were determined to a precision of 4\%.)
The parity-violating asymmetry was measured using higher
beam currents, for which it was necessary to integrate the detector signals
over the beam pulse. 
The incident electron beam was pulsed at 600 Hz; 
the signals from the detector, beam toroid monitors, and various other beam 
monitors were integrated and digitized for every 25 $\mu$sec long 
beam pulse. The parity-violating asymmetry $A$ was determined from the 
asymmetries in ratios of integrated detector signal to 
beam intensity for left- and right-handed beam pulses.

The polarized electron beam was generated 
via photoemission from unstrained 
GaAs by polarized laser light. The laser beam helicity
for each pulse was determined by a $\lambda /4$ Pockels cell and
was randomly chosen for each of 10 consecutive beam pulses;
the complement helicities were then used for the next 10 pulses. 
The asymmetry in the normalized detector yields 
was computed for ``pulse pairs'' separated by 1/60 of a second to 
minimize systematic errors. 
The electron beam helicity relative to all electronic signals
can be manually reversed by inserting a $\lambda$/2 plate in the laser beam. 
(We denote this
configuration as $\lambda$/2 = ``IN'' as opposed to $\lambda$/2 =
``OUT''.) During the 1998 running period, the IN/OUT configuration 
was reversed every few days to minimize false asymmetries
and test for systematic errors. 
The electron polarization was measured using a M{\o}ller system on the
beamline and averaged 36.3$\pm$1.4 \% during the experiment. 
The effect of small transverse components of
electron polarization on the observed parity violation signal was 
studied and determined to be negligible.

Helicity correlations of various parameters of the
electron beam were monitored continuously during the experiment. These
parameters include the beam intensity,
position and angle at the target in both transverse
dimensions ($x$ and $y$), the beam energy, and the beam ``halo''.
Two forward angle lucite {\v C}erenkov
counters were also implemented 
at $\sim 12^\circ$ to monitor luminosity and test for 
helicity dependence. These monitors detected low $Q^2$
elastic scattering at forward angles and other soft 
electromagnetic radiation and should
show negligible parity violating asymmetry.

As in the past, we reduced the beam intensity asymmetry through
an active feedback system. In 1998, this feedback was implemented
with an additional Pockels cell located between linear 
polarizers to separate this function from the Pockels cell that
controlled the helicity (HPC). The HPC was also repositioned 
to be downstream of all laser transport elements. These resulted in 
improved stability of the
laser beam position under helicity reversal. In addition, we
implemented a feedback system to reduce the remaining helicity-correlated
beam position asymmetry\cite{Aver99}. This was accomplished using a tilted
glass plate in the laser beam path and a piezoelectric
transducer. By adjusting the tilt of this glass plate with helicity
reversal, the first order beam position asymmetry is
reduced, resulting in improved quality of the data. For example, 
the helicity correlated vertical beam shift at the target
was reduced from $\sim 200$ nm to typically $< 20$ nm.

Of the 110~C of beam delivered to the experiment in 1998, the first
24~C were taken before the position feedback system was
fully implemented, and significant position asymmetries were
present. This is evident in Figure 1a, where the luminosity monitor
asymmetries are shown for this time frame 
(``piezo off'') in comparison with the later runs (``piezo on'').
We also use a linear regression technique to remove such
effects from the data~\cite{Muel}. 
The results of this analysis are shown in Figure 1b,
where the corrected asymmetries are displayed. This procedure 
(involving six beam parameters: $x$, $y$, $\theta_x$, $\theta_y$, energy,
and intensity) is very effective at removing the effects of beam
helicity correlations, resulting in a final corrected 
luminosity monitor asymmetry result of $0.17 \pm 0.11$ ppm. 
In Figure 2 are shown the analogous plots for the asymmetry measured in the
SAMPLE detector (all 10 mirrors combined
and corrected for background dilution, radiative effects,
and beam polarization). In contrast to the luminosity monitors,
the detector asymmetry is quite robust with respect to beam helicity
correlations, the corrections affecting
the final asymmetry by only 5\% or 0.2 ppm. This correction
is about equal to the estimated systematic error in the procedure as
determined from the luminosity monitor analysis.

The elastic scattering asymmetry was determined from the 10 individual mirror
asymmetries after correction for all effects, including background
dilution. The measured shutter closed asymmetry for all 10 mirrors combined 
(appropriately scaled to compare directly to the elastic asymmetry),
is $-0.57 \pm 0.64$ ppm,
consistent with zero as expected assuming the shutter closed
yield is dominated by low-$Q^2$ processes. 
However, the mirror-by-mirror distribution of 
shutter closed asymmetries is statistically improbable,
indicating the presence of some non-statistical component
to the shutter closed yield. We therefore assume the combined shutter 
closed asymmetry to be zero in our analysis, and assign a 
systematic error due to the uncertainty in the 
shutter closed asymmetry of 0.64 ppm. 

The resulting elastic asymmetry is
\begin{equation}
A = -4.92 \pm 0.61 \pm 0.73 \> {\rm ppm} \label{eq:1}
\end{equation}
where the first uncertainty is statistical and the second is the 
estimated systematic error as summarized in Table 1. 
This value is in good agreement with our previous
reported measurement~\cite{Muel97}.

At the mean kinematics of the experiment ($Q^2$ = 0.1 (GeV/c)$^2$
and $\theta$=146.1$^\circ$), the theoretical asymmetry is
\begin{equation}
A = -5.61 + 3.49 G_M^s + 1.55 G_A^Z\label{eq:3}
\end{equation}
where
\begin{equation}
G_A^Z = -(1+R_A^1)G_A + R_A^0 + G_A^s\> .\label{eq:2}
\end{equation}
$G_A$ is the charged current nucleon form factor: we
use $G_A=G_A(0)/(1+{Q^2\over M_A^2})^2$, with 
$G_A(0)=-(g_A/g_V) = 1.267\pm 0.035$~\cite{Cas98} and 
$M_A=1.061\pm 0.026$~(GeV/c)
\cite{Gar93}. $G_A^s(Q^2=0)=\Delta s=-0.12\pm 0.03$~\cite{Lip99},
and $R_A^{0,1}$ are the isoscalar and isovector
axial radiative corrections. The radiative
corrections were estimated by ref.~\cite{Mus90} to be
$R_A^1=-0.34$ and $R_A^0=-0.12$, but with nearly 
100\% uncertainty.\footnote{The notation used here is
$R_A^0=(1/2)(3F-D)R_A^{T=0}$, where $\sqrt{3}R_A^{T=0}=-0.62$
in ref.~2b} 

The strange magnetic form factor derived
from the asymmetry in Eq. \ref{eq:1} is
\begin{eqnarray}
G_M^s (Q^2 = 0.1 {\rm (GeV/c)}^2) = +0.197 &\pm& 0.17 \pm 0.21 \\
&-&  0.445 G_A^Z  \> 
{\rm n. m.} \nonumber
\end{eqnarray}
This result is graphically displayed in Fig.~3 along with 
$G_A^Z$ (dashed line) computed from taking 
the above calculated values of $R_A^0$ and $R_A^1$. 
Combining this value of $G_A^Z$ with our 
measurement implies a substantially positive value
of $G_M^s (Q^2 = 0.1 {\rm (GeV/c)}^2) = +0.61 \pm 0.17 \pm 0.21$.
As noted in recent papers~\cite{Bei96,Don98} most model calculations 
tend to produce negative values of $\mu_s = G_M^s(Q^2 = 0)$, 
typically about $-0.3$. A recent calculation using 
lattice QCD techniques (in the quenched approximation) reports a result 
$\mu_s = -0.36 \pm 0.20$~\cite{Don98}. As shown in Fig.~3, 
our new measurement implies that the computed
negative value of $G_A^Z$ is inconsistent with $G_M^s < 0$.
Another recent study
using a constrained Skyrme-model Hamiltonian that fits the
baryon magnetic moments yields a positive value of 
$\mu_s = +0.37$~\cite{Hon97},
which is in better agreement with our measurement and with 
the calculated value of $G_A^Z$.

Since the dominant uncertainty in $G_A^Z$ comes from
$R_A^1$, eliminating the uncertainty in $R_A^1$ is essential for
deriving a firm conclusion about $G_M^s$. Toward this end,
we are presently running the SAMPLE experiment with a deuterium 
target to measure the quasielastic asymmetry from deuterium~\cite{Pitt94}. 
This asymmetry is quite insensitive to strange quark effects and 
will therefore independently
determine the isovector axial radiative correction. 
When the deuterium data are available, we will be able to
provide definitive experimental information on the proton's
strange magnetic form factor. 

The skillful efforts of the staff of the MIT/Bates facility to provide
high quality beam and improve the experiment are gratefully acknowledged.

This work was supported by
NSF grants PHY-9420470 (Caltech), PHY-9420787 (Illinois), 
PHY-9457906/PHY-9971819 (Maryland), PHY-9733772 (VPI) and DOE
cooperative agreement DE-FC02-94ER40818 (MIT-Bates) and
contract W-31-109-ENG-38 (ANL).


\begin{figure}
\centerline{\psfig{figure=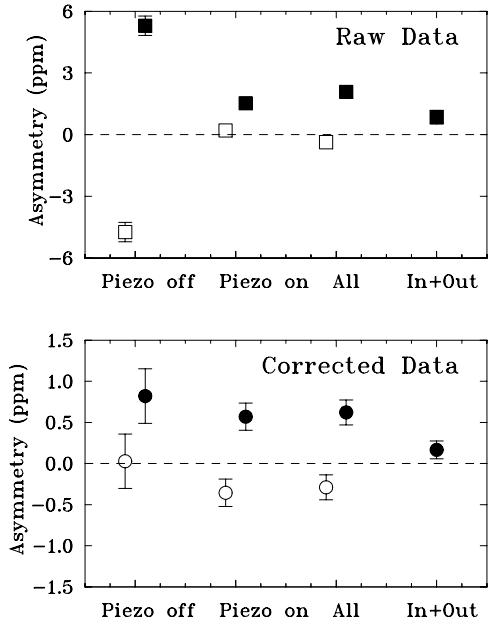,width=2.5in}}
\vskip 0.2 in
\caption{Measured asymmetries in the luminosity monitors in
parts per million (ppm). The open symbols are for $\lambda/2=$IN and
the filled symbols are for $\lambda/2=$OUT. The upper plot shows the
raw data and the lower plot are the data corrected for beam
parameter asymmetries. The beam polarization is accounted for and the
error bars include
statistical errors only. Note the change in scale in the lower plot.}
\end{figure}

\begin{figure}
\centerline{\psfig{figure=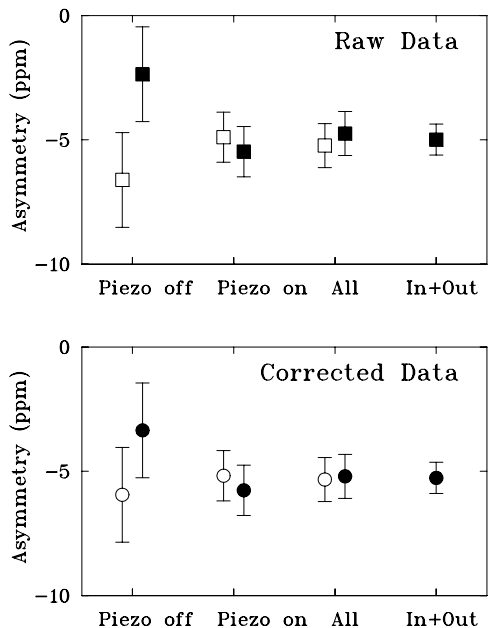,width=2.5in}}
\vskip 0.2 in
\caption{Measured asymmetries for the combined {\v C}erenkov detectors.
The open symbols are for $\lambda/2=$IN and
the filled symbols are for $\lambda/2=$OUT. The upper plot shows the
raw data and the lower plot are the data corrected for beam
parameter asymmetries. The
error bars are statistical only. Including (mirror-by-mirror) the systematic 
error changes the relative weights of each mirror, resulting in the
slightly smaller value of $A$ in equation 2.}
\end{figure}

\begin{figure}
\centerline{\psfig{figure=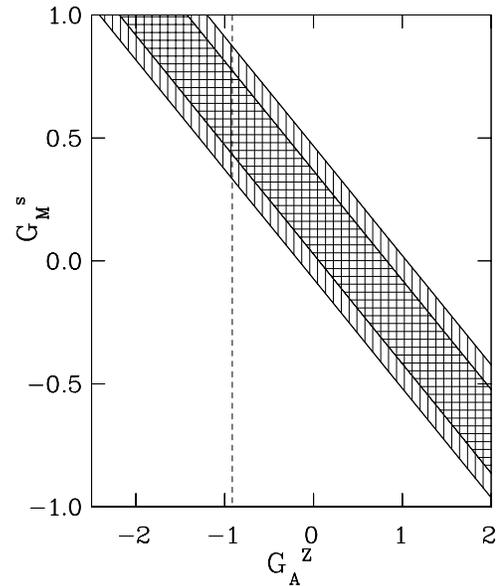,width=2.5in}}
\vskip 0.2 in
\caption{Error band of $G_M^s$ for the allowed region (shaded) corresponding 
to the present measurement of the parity-violating asymmetry. 
The inner hatched region includes the statistical error, the outer 
represents the systematic uncertainty added in quadrature, and 
the vertical dashed line corresponds to the calculated value of $G_A^Z$
(as defined in eq.~4) using theoretical estimates of ref.~2 for 
$R_A^0$ and $R_A^1$.}
\end{figure}

\begin{table}
\begin{center} 
\caption{Summary of relative uncertainties on the measured asymmetry.}
\begin{tabular}{l|c|} \hline
 Source & $\delta$A/A (\%) \\ \hline
 Shutter closed asymmetries & 13 \\
 Helicity-correlated corrections procedure & 5 \\
 Unpolarized background dilution factor & 4 \\
 Beam polarization determination & 4 \\ 
 Nucleon electromagnetic form factors & 3 \\ \hline
 Total systematic (added in quadrature) & 15 \\ 
\hline
\end{tabular}
\end{center}
\end{table}


\end{multicols}
\end{document}